%% file: reciprocaltaucell.tex
\documentclass[conference]{IEEEtran}
\usepackage{amsmath,amssymb}
\usepackage{graphicx}
\usepackage{booktabs}
\usepackage{siunitx}
\usepackage{cite}
\usepackage[hidelinks]{hyperref}
\usepackage{xcolor}
\usepackage{circuitikz}
\usetikzlibrary{calc,arrows.meta}
\definecolor{tblue}{HTML}{2A78D6}
\definecolor{torange}{HTML}{EB6834}
\definecolor{idl}{HTML}{C07A00}
\newcommand{\T}{^{\!\top}}
\newcommand{\A}{\mathcal{A}}
\newcommand{\purpose}{\emph{Purpose.}~}
\newcommand{\setup}{\emph{Setup.}~}
\newcommand{\finding}{\emph{Finding.}~}

\begin{document}
\bstctlcite{IEEEexample:BSTcontrol}

\title{The Reciprocal $\tau$-Cell: A Log-Domain\\Circuit  for On-Chip Learning}

\author{\IEEEauthorblockN{Craig Jin and Thomas Chaffey}
\IEEEauthorblockA{School of Electrical and Computer Engineering, The University of Sydney, NSW 2006, Australia\\
craig.jin@sydney.edu.au, thomas.chaffey@sydney.edu.au}}

\maketitle

\begin{abstract}
Training hardware by gradient descent requires a gradient to be computed using a separate backwards pass.
If a circuit is \emph{reciprocal}, meaning every connection is symmetric, the backward pass is the same
network run a second time with the error as its input. The \emph{reciprocal $\tau$-cell}, a
weak-inversion log-domain circuit, makes this practical in standard CMOS. Each cell is a
current-mode leaky integrator with a built-in ReLU; each connection is two transistors whose signed
strength is a ratio of bias currents. The equilibrium of the network 
is the solution to a convex program, and latching the cells that are off will \emph{hardware linearise} the circuit, making its second run exactly
linear. Transistor-level SKY130 simulations show that the second run gives the gradient within
1.9\%; that hardware linearisation permits large, noise-robust error signals; that a 13-cell network learns the Iris data set; and that learning in situ restores mismatched circuits on which software-trained weights
fail, from 53--63\% to 88--96\% test accuracy.
\end{abstract}


%======================================================================
\section{Introduction}
%======================================================================

A neural network is trained using gradient descent. Computing the gradient is expensive and difficult to implement in hardware: digital processors run a separate backward pass~\cite{hajizada_online_2026}, analog
in-memory chips compute it in digital periphery~\cite{zhang_edge_2023,athena_demonstration_2024}, and
log-domain designs use dedicated update circuits~\cite{tageldeen_learning_2025}.

There is a more elegant route. In a network that settles to an equilibrium, the backward
pass is a settling problem on the transposed network. If every connection is symmetric, that
\emph{is} the original network: the backward pass is a second run of the same circuit with the
error as input. This principle underlies contrastive learning \cite{Hinton1989, Movellan1991, Huijzer2026}, equilibrium
propagation~\cite{scellier_equilibrium_2017,kendall_training_2020}, coupled learning in resistor and
transistor networks~\cite{dillavou_demonstration_2022,dillavou_machine_2024}, and the Boltzmann-machine
chips of Alspector \emph{et al.}~\cite{alspector_stochastic_1987,alspector_performance_1988,alspector_experimental_1991}.
Watfa \emph{et al.} show with the adjoint method that the precise condition is reciprocity of the
circuit's small-signal (or linearized) model~\cite{watfa_designing_2025}, a property first studied in the context of automatic circuit design by Director and Rohrer \cite{Director1969} and recently shown by B\"osch \emph{et al.} to unify a wide range of in-situ learning algorithms~\cite{bosch_unifying_2026}. 
Chaffey relates reciprocal resistor--diode
networks to a particular class of machine learning models,  ReLU monotone operator equilibrium networks (monDEQ) \cite{chaffey_circuit_2026}.

These circuits, summarised in Table~\ref{tab:prior}, are passive resistor or memristor networks: negative weights need duplicated
layers and there are no programmable dynamics. Above all, because the adjoint method requires the small-signal model of the circuit and the circuit is nonlinear,
``the injected currents must be kept as small as possible''~\cite{watfa_designing_2025, bosch_unifying_2026}, and small
signals are buried in noise~\cite{watfa_end--end_2026,oh_memristor_2023}.
Chaffey introduces an alternative method to deal with nonlinearity, replacing a nonlinear circuit directly with its small-signal model through a process called 
\emph{hardware linearisation}~\cite{chaffey_circuit_2026}, which fixes diodes in their conducting/non-conducting state.

This paper presents the \textbf{reciprocal $\tau$-cell}, which implements hardware linearisation in standard
CMOS. It builds on the $\tau$-cell, a log-domain first-order section for arbitrary differential
equations~\cite{van_schaik_tau-cell_2003}, and adds a two-transistor connection whose signed strength
is a ratio of bias currents, identical in both directions by construction, and a latched backward
run that is exactly linear for any error amplitude. Section~\ref{sec:results} tests, in
transistor-level simulations, that (i)~the cell and connection behave as designed, (ii)~the second
run gives the gradient, (iii)~latching allows large, noise-robust error signals, (iv)~mismatch
degrades the gradient gracefully, (v)~the circuit learns a benchmark task as well as software, and
(vi)~learning in situ, on the circuit itself with its own measured gradient, repairs mismatch that
defeats weights trained in software and copied in.

\begin{figure*}[t]
\centering
\resizebox{\textwidth}{!}{\input{fig1_circuit.tex}}
\caption{The reciprocal $\tau$-cell. (a)~One cell with the circuitry that operates it. $M_1$
rectifies and takes the logarithm of the summed current at $A_i$; the translinear loop
$M_1$--$M_4$ with $C$ gives $\tau\dot x_i=s_i-x_i$. Switches inject the input ($\phi_1$) or the
pedestal plus error ($\phi_2$); a comparator latches whether the cell is on ($a_i$), and $S_L$ ties
$A_i$ to $V_{\rm ref}$ in cells that are off; $\phi_{\rm rd}$ reads $x_i$ and $\phi_h$ holds it.
(b)~One reciprocal connection $i\!\leftarrow\!j$: $M_d$, biased by $I_w$, shifts the sender's
capacitor voltage, and $M_w$ delivers $(I_w/I_0)x_j$ out of $A_i$ (negative weight) or, through a
PMOS mirror, into it (positive). One code sets $I_w$ for both directions. Black: simulated at
transistor level. Orange: ideal in simulation.}
\label{fig:schematic}
\end{figure*}
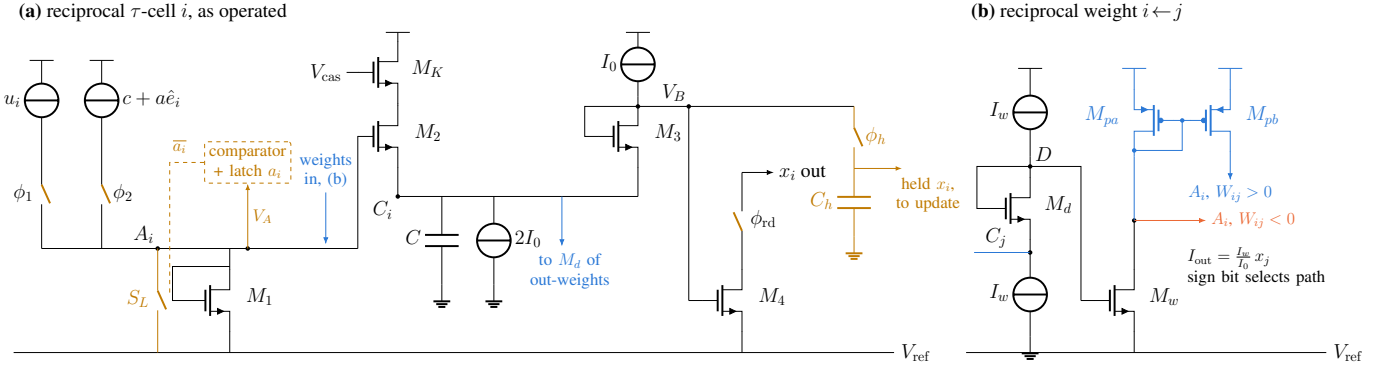

%======================================================================
\section{What the Network Computes}
\label{sec:what}
%======================================================================

\subsection{Equilibrium behavior}
Each cell adds its input current $u_i$ to the currents its neighbours send it, rectifies the
sum, and relaxes towards it with time constant $\tau$:
\begin{equation}
\tau\frac{dx_i}{dt}=-x_i+\mathrm{ReLU}\Big(u_i+\sum_jW_{ij}x_j\Big).
\label{eq:dyn}
\end{equation}
Here $x_i$ is the cell's output current and $\mathrm{ReLU}(s)=\max(0,s)$ is the rectified linear
unit. The networks settles to the equilibrium 
\begin{equation}
x=\mathrm{ReLU}(Wx+u).
\label{eq:fixed}
\end{equation}
If the connections are symmetric, $W=W\T$, and feedback is not too strong, $I-W\succ0$, the
equilibrium is unique and reached from any starting point~\cite{winston_monotone_2020}. It is
then the solution of a convex optimisation problem:
\begin{equation}
x=\arg\min_{x\ge0}\ \tfrac12x\T(I-W)x-u\T x .
\label{eq:qp}
\end{equation}
So one circuit serves two purposes. Without learning, it solves non-negative quadratic programs
(non-negative least squares, sparse coding~\cite{rozell_sparse_2008}, winner-take-all), in the
tradition of Hopfield and Tank~\cite{hopfield_neural_1985}. With learning, it realizes a monDEQ: it
behaves like an infinitely deep network with tied weights, evaluated in one settling
time~\cite{winston_monotone_2020,chaffey_circuit_2026}.

\subsection{Gradient computation}
Training adjusts $W$ to reduce a loss $\mathcal{L}(x)$. Take two cells, both on, coupled by $w$
each way: $x_1=u_1+wx_2$, $x_2=u_2+wx_1$. Extra current into cell~1 changes $x_2$ exactly as much as
extra current into cell~2 changes $x_1$,
$\partial x_2/\partial u_1=\partial x_1/\partial u_2=w/(1-w^2)$: the network is
\emph{reciprocal}. Now let cell 2 be the output with error $e_2=x_2-y$. Differentiating the
equilibrium gives $\partial\mathcal{L}/\partial w=e_2(x_1+wx_2)/(1-w^2)$. If we instead remove the
inputs, inject $e_2$ into cell~2 and let the same circuit settle, the cells settle to
$d_1=we_2/(1-w^2)$ and $d_2=e_2/(1-w^2)$, and $d_1x_2+d_2x_1$ is exactly that gradient.

The same holds for any number of cells. Let $D$ be the diagonal matrix of which cells are on
($D_{ii}=1$) or off ($D_{ii}=0$). At fixed $D$, $x=(I-DW)^{-1}Du$, and for a shared weight
$W_{ij}=W_{ji}$
\begin{equation}
\frac{\partial\mathcal{L}}{\partial W_{ij}}=d_ix_j+d_jx_i,\qquad d=D(I-W\T D)^{-1}e,
\label{eq:grad}
\end{equation}
with $e=\partial\mathcal{L}/\partial x$. The Jacobian $J=I-DW$ is \emph{not} symmetric even when
$W$ is, but it is \emph{symmetrisable}: where $D$ is invertible, $J\T=D^{-1}JD$. For a ReLU, $D$ has
zeros; removing the cells that are off leaves the symmetric $I-W_{\A\A}$ on the active set $\A$, and
$d=(I-DW)^{-1}De$ is \emph{the response of the same circuit, with the same cells on and off, to the
error injected as an input}. Each weight's gradient uses currents at its own two cells only.

%======================================================================
\section{The Circuit}
\label{sec:circuit}
%======================================================================

\begin{figure*}[t]
\centering
\includegraphics[width=\textwidth]{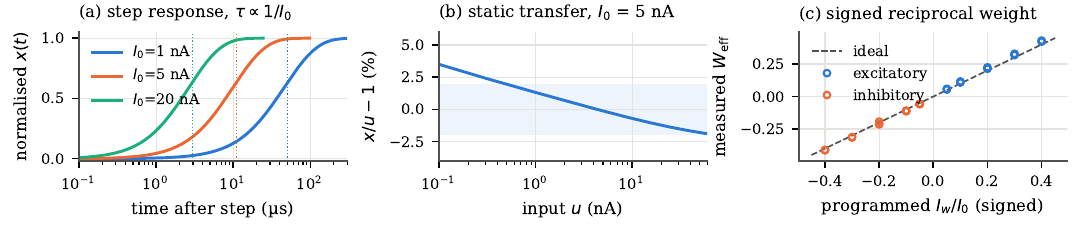}
\caption{Do the building blocks work? (a)~Step response of one cell at three bias currents
$I_0$; dotted: measured $\tau$. (b)~Static error of one cell, $x/u-1$, over an input sweep
($\pm2\%$ band). (c)~Weight of a reciprocal pair measured against the programmed value, both signs,
three operating points each.}
\label{fig:circuit}
\end{figure*}

\subsection{Operation of a cell}
In weak inversion, with its body tied to its source (each NMOS in its own deep n-well), a MOS
transistor obeys $I=I_S\exp\big((V_G-V_S)/nU_T\big)$, $n\approx1.5$: every gate--source voltage is
the logarithm of a current, and around a loop of them products of currents balance (the
translinear principle~\cite{gilbert_translinear_1996}). In one cell (Fig.~\ref{fig:schematic}(a)):
\emph{Summing node $A_i$:} the input $u_i$ and all neighbour currents add, $s_i=u_i+\sum_jW_{ij}x_j$.
\emph{$M_1$, ReLU and logarithm:} the sum flows into the diode-connected $M_1$, so
$V_A=V_{\rm ref}+nU_T\ln(s_i/I_S)$. A diode realises a RELU: if the sum is negative, $M_1$
turns off and so does the cell.
\emph{$M_2$, $M_3$, $M_4$ and $C$:} $M_3$ carries a bias $I_0$ and sits on the capacitor node, and
$M_4$ copies its gate voltage to the output, $x_i=I_0\exp\big((V_C-V_{\rm ref})/nU_T\big)$. The
loop $M_1$--$M_4$ forces the current that $M_2$ delivers to the capacitor to be $I_2=s_iI_0/x_i$.
With the $2I_0$ sink, $C\dot V_C=I_2-I_0$, and since $\dot x_i=(x_i/nU_T)\dot V_C$,
\begin{equation}
\tau\frac{dx_i}{dt}=s_i-x_i,\qquad \tau=\frac{nCU_T}{I_0},
\label{eq:cell}
\end{equation}
which is one row of~\eqref{eq:dyn}: exactly linear in current while on, with a time constant set by
a bias current. The cascode $M_K$ suppresses the Early effect of $M_2$.

\begin{figure}[t]
\centering
\includegraphics[width=0.92\columnwidth]{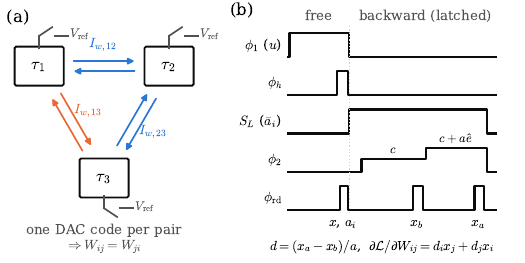}
\caption{(a)~Network; each pair shares one weight code. (b)~One learning step with the switches
of Fig.~\ref{fig:schematic}(a): free run, hold $x$, latch $a_i$, two backward measurements.}
\label{fig:network}
\end{figure}

\subsection{A weight is a voltage shift}
Two transistors are used for a connection between two cells. They hold the logarithm of each cell's output on a wire (Fig.~\ref{fig:schematic}(b)). A diode $M_d$, biased by a static current $I_w$, sits on
the sender's capacitor node, so its gate sits at $V_{C_j}+nU_T\ln(I_w/I_S)$. Adding that voltage
multiplies the current: $M_w$, with its source at $V_{\rm ref}$, conducts
\begin{equation}
I_{\rm out}=\frac{S_w}{S_d}\frac{I_w}{I_0}\,x_j\equiv|W_{ij}|\,x_j .
\label{eq:weight}
\end{equation}
A ratio of bias currents is set by a current DAC or bias generator~\cite{delbruck_bias_2005}; an
equal sink returns $I_w$ to $C_j$. A negative weight draws its current out of $A_i$; a positive
weight is pushed in by a PMOS mirror. Both arrive \emph{before} $M_1$, so the ReLU acts on the total,
as~\eqref{eq:fixed} requires. One code drives both directions (Fig.~\ref{fig:network}(a)), so $W=W\T$
holds by construction rather than by trimming.

\subsection{Hardware linearisation}
The second run must remove the cells that were off. After the free run, the comparator latches
$a_i$ and $S_L$ ties the summing node of each cell that was off to $V_{\rm ref}$, silencing it.
Closing every $S_L$ at power-up also avoids a second, high-current equilibrium. The remaining cells
are exactly linear. Because currents cannot be negative while errors can, the inputs are replaced by
a pedestal $c$ on the active cells, and the outputs are measured with and without the error added
(Fig.~\ref{fig:network}(b)):
\begin{equation}
d=\frac{x(c\mathbf{1}+a\hat e)-x(c\mathbf{1})}{a}=(I-W_{\A\A})^{-1}\hat e .
\label{eq:adj}
\end{equation}
This is exact for \emph{any} amplitude $a$ that keeps the active cells on: the circuit realisation
of hardware linearization~\cite{chaffey_circuit_2026}. Each weight is then updated by
$-\eta(d_ix_j+d_jx_i)$.

\begin{figure*}[t]
\centering
\includegraphics[width=\textwidth]{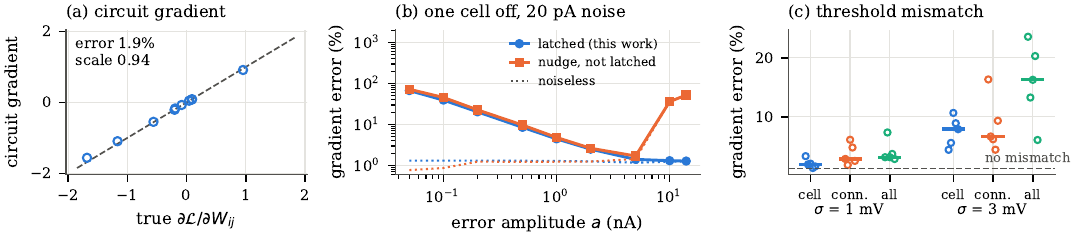}
\caption{Does the circuit measure its own gradient? Six-cell networks.
(a)~Circuit-measured against true gradient of all ten weight pairs (Sec.~\ref{sec:grad}).
(b)~Gradient error as the error amplitude $a$ grows, with one cell off and \SI{20}{pA} readout noise;
dotted: no noise (Sec.~\ref{sec:latch}). (c)~Gradient error with random threshold offsets on cell,
connection or all transistors; 5 draws each, bar: median, dashed: no mismatch (Sec.~\ref{sec:mm}).}
\label{fig:grad}
\end{figure*}

%======================================================================
\section{Simulation Results}
\label{sec:results}
%======================================================================

\subsection{Simulation setup}
\label{sec:method}
Six questions are answered below, one per subsection, by transistor-level ngspice simulations
with the SKY130 models (typical corner, \SI{27}{\celsius}):
$W/L=\SI{4}{\micro m}/\SI{8}{\micro m}$, $C=\SI{1}{pF}$, $I_0=\SI{5}{nA}$, \SI{1.8}{V} supply. Every transistor drawn black in Fig.~\ref{fig:schematic} is simulated; the orange blocks, the
current sources and the update arithmetic are ideal, and each phase of Fig.~\ref{fig:network}(b) is
solved as a settled (DC) operating point. A gradient is judged against
the \emph{true} gradient of the simulated circuit, found by raising and lowering each shared weight
slightly and re-solving the circuit (central finite differences). The \emph{gradient error} is the relative error of the circuit gradient $g_c$ against the true
gradient $g_t$ after the best uniform rescaling, $\min_s\lVert sg_c-g_t\rVert/\lVert g_t\rVert$,
given in percent; a uniform scale is harmless because the learning rate absorbs it. Mismatch is a Monte
Carlo of random threshold offsets on the transistors; for these devices SKY130 predicts about
\SI{1.3}{mV}.

\subsection{Do the cell and the connection work as designed?}
\purpose Check~\eqref{eq:cell} and~\eqref{eq:weight} before building networks.
\setup One cell is driven by input steps and a slow input sweep; the weight between two coupled
cells is extracted from their responses to small input changes, independently of each cell's gain.
\finding The cell is first order with $\tau\propto1/I_0$: 51, 11 and \SI{2.9}{\micro s} at $I_0=1$,
5 and \SI{20}{nA} (Fig.~\ref{fig:circuit}(a)), about 1.3 times $nCU_T/I_0$ because transistor
capacitances add to $C$. It copies its input within $\pm2\%$ from 0.5 to \SI{60}{nA}
(Fig.~\ref{fig:circuit}(b)). Programmed weights from $\pm0.05$ to $\pm0.4$ are reproduced within
0.015 when negative and 0.03 when positive, where the uncascoded PMOS mirror adds an Early effect
excess (Fig.~\ref{fig:circuit}(c)). Without learning, six mutually inhibiting cells settle from a power-up reset to the solution
of~\eqref{eq:qp} in \SI{0.12}{ms}, with the right three cells on and each current within 1\%. Six
cells with ten pairs draw \SI{0.26}{\micro W}.

\subsection{Does the second run give the gradient?}
\label{sec:grad}
\purpose Test the central claim,~\eqref{eq:grad} with~\eqref{eq:adj}.
\setup Six cells joined by ten pairs of mixed sign, all cells on, two of them outputs with loss
$\mathcal{L}=\tfrac12\sum_k\big((x_k-y_k)/I_0\big)^2$ and targets away from the outputs. The circuit
runs freely, then twice latched, and forms $d_ix_j+d_jx_i$ for every pair.
\finding All ten gradients agree with the true ones (Fig.~\ref{fig:grad}(a)): the gradient error is
1.9\% and the scale 0.94. The residual comes from the small cell and weight errors of
Fig.~\ref{fig:circuit}.

\subsection{Why hardware linearise?}
\label{sec:latch}
\purpose A readout is noisy, so the error signal should be large, but in a nonlinear circuit a large
error can switch cells on or off.
\setup The same network with weights that switch one cell off; \SI{20}{pA} rms of readout noise is
added to every measured current (200 trials per point). The latched run is compared, over error amplitude, with a
nudge from the free state as in equilibrium propagation and~\cite{watfa_designing_2025}.
\finding At small amplitudes both are limited by noise (Fig.~\ref{fig:grad}(b)). Latched, the gradient error
keeps falling, to 1.3\% at \SI{14}{nA}. Unlatched, the nudge is best at \SI{5}{nA} (1.7\%) and fails at 10 and
\SI{14}{nA} (36\% and 52\%), because the error switches the off cell on. Latching removes the
trade-off between noise and linearity.

\subsection{How much does mismatch matter?}
\label{sec:mm}
\purpose Real transistors differ.
\setup The network of Sec.~\ref{sec:latch} with random threshold offsets on the cell transistors, the
connection transistors, or all, at 1 and \SI{3}{mV}; each draw is compared with the true gradient of
its own mismatched circuit.
\finding At \SI{1}{mV} the median gradient error rises only from 1.3\% to 3.1\%; at \SI{3}{mV} it is
7.9\%, 6.7\% and 16\% (Fig.~\ref{fig:grad}(c)). A cell
offset changes that cell's gain, turning $W$ into $GW$, which is diagonally symmetrisable: the second
run stays exact up to a per-cell factor on $d$, which could be calibrated out. A connection offset
breaks $W_{ij}=W_{ji}$, which no rescaling restores in a loop of three or more cells.

\begin{figure}[t]
\centering
\includegraphics[width=\columnwidth]{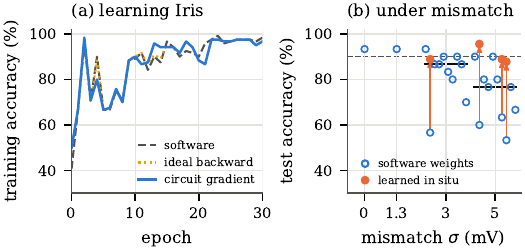}
\caption{Learning a benchmark, Iris. (a)~Training accuracy per epoch: circuit learning on its own
gradient, circuit with an ideal backward pass (15 epochs), and software (Sec.~\ref{sec:iris}).
(b)~Test accuracy of software-trained weights copied into circuits with threshold mismatch, one
open dot per draw (bar: median of ten; dashed: software), and of the four worst after learning in
situ (filled) (Sec.~\ref{sec:insitu}).}
\label{fig:iris}
\end{figure}

\subsection{Does the circuit learn a real task?}
\label{sec:iris}
\purpose Go from one gradient to learning.
\setup The Iris data set: 150 flowers of three species, four measurements each; 120 for training,
30 for testing. Four input cells receive the measurements as \SIrange{1}{12}{nA} and connect to six
hidden cells, which connect to three output cells, one per species: 42 pairs with 6-bit signed
weights, a signed bias on each hidden and output cell, and about 600 transistors. The target is
\SI{6}{nA} on the output of the true species and \SI{3}{nA} on the others; the prediction is the largest
output. For each flower the circuit runs once freely and twice latched, and every weight moves by
$-\eta(d_ix_j+d_jx_i)$, $\eta=0.02$.
\finding The circuit follows the training curve of the same network in software, with the exact
gradient, epoch by epoch (Fig.~\ref{fig:iris}(a)). After 30 epochs it reaches 96.7\% training and
90.0\% test accuracy, against 98.3\% and 90.0\%, with the same three test errors. 
%On the flowers that drive learning, its gradient is within 5--8\% of the
%true one.

\subsection{Why learn on chip?}
\label{sec:insitu}
\purpose The usual alternative is to train in software and copy the weights.
\setup The software weights are copied into circuits with 1.3, 3 and \SI{5}{mV} of threshold mismatch
(ten draws each at 3 and \SI{5}{mV}, standing for smaller transistors). The four worst are then
retrained on their own circuit, from the software weights, with only the circuit's gradient.
\finding At \SI{1.3}{mV} the copied weights keep 93.3\% test accuracy, but the median falls to
86.7\% at \SI{3}{mV} and 76.7\% at \SI{5}{mV}, and to 53\% in the worst draw (Fig.~\ref{fig:iris}(b)),
partly because output-cell gains rescale the class scores, differently on every chip. Learning in
situ, with both runs and so the gradient measured on the mismatched circuit itself, raises the four
worst draws from 53--63\% to 88--96\% (mean of the last three epochs). 
%despite the fact that at \SI{5}{mV} the computed 
%gradient is about 35\% off the true one.

\begin{table}[t]
\caption{Analog circuits that compute their own gradient}
\label{tab:prior}
\centering
\scriptsize
\setlength{\tabcolsep}{2.4pt}
\begin{tabular}{@{}lllll@{}}
\toprule
Ref. & Circuit & Signed $W$ & Backward pass & Validation\\
\midrule
\cite{alspector_performance_1988,alspector_experimental_1991} & CMOS, binary units & sign + 4\,b & correlation, two phases & chip\\
\cite{kendall_training_2020,oh_memristor_2023} & resistor / memristor & duplicated & EP, small nudge & SPICE\\
\cite{dillavou_demonstration_2022,dillavou_machine_2024} & resistor / MOS & -- & coupled, twin network & bench\\
\cite{watfa_designing_2025,watfa_end--end_2026} & R, op-amp, memristor & duplicated & adjoint, small/finite & SPICE\\
\cite{zhang_edge_2023,athena_demonstration_2024} & memristor crossbar & differential & transposed read, digital & chip\\
\cite{chaffey_circuit_2026} & passive R--diode & duplicated & HW linearization, exact & ngspice\\
This work & log-domain CMOS & native ($I_w$) & latched, exact & SKY130, Iris\\
\bottomrule
\end{tabular}
\end{table}

%======================================================================
\section{Conclusion}
\label{sec:discussion}
%======================================================================

The reciprocal $\tau$-cell allows a monDEQ \cite{winston_monotone_2020} to be both implemented and trained in a CMOS circuit, using hardware linearisation \cite{chaffey_circuit_2026} to compute the gradient with the circuit itself.
%In Table~\ref{tab:prior}, memristor chips are larger but form the gradient
%digitally~\cite{zhang_edge_2023,athena_demonstration_2024}; equilibrium-propagation circuits reach
%MNIST in simulation with duplicated layers and nudges~\cite{watfa_designing_2025,watfa_end--end_2026}.
We target low power edge applications such as learning at the sensor. A current DAC per pair
dominates area, and the products $d_ix_j$, here formed numerically, need a translinear multiplier per
weight~\cite{alspector_performance_1988}. 
%Cell pairs give signed variables and linearly constrained
%programs~\cite{chaffey_circuit_2026}; an $I_w$ driven by another cell is a fast weight, the ingredient
%of linear-attention models~\cite{schlag_linear_2021}.  
In this work we learn static mappings between vectors; learning dynamic sequence mappings in situ remains an open problem~\cite{bosch_unifying_2026,scurria_equilibrium_2026}.

\bibliographystyle{IEEEtran}
\bibliography{IEEEcontrol,references,Tom_refs}

\end{document}

%% file: fig1_circuit.tex
% Fig. 1 of the ISCAS paper: the reciprocal tau-cell as operated (circuitikz).
% Black: simulated at transistor level in SKY130.  Orange: ideal in simulation.
\begin{circuitikz}[font=\Large, line width=0.7pt,
  blk/.style={draw=idl, dashed, rounded corners=2pt, align=center, font=\large, text=idl}]
\ctikzset{tripoles/mos style/arrows}
%=========================== (a) operating cell ===========================
\node[anchor=west] at (-7.0,4.6) {\textbf{(a)} reciprocal $\tau$-cell $i$, as operated};
\draw (-7.0,-3.9) -- (15.0,-3.9); \node[right] at (15.05,-3.9) {$V_{\rm ref}$};
% ---- switched input and adjoint injection (ideal switches) ----
\draw (-6.6,3.4) -- (-6.0,3.4);
\draw (-6.3,3.4) to[isource,l_=$u_i$] (-6.3,1.4) to[nos,l_=$\phi_1$,color=idl] (-6.3,-1.3);
\draw (-5.1,3.4) -- (-4.5,3.4);
\draw (-4.8,3.4) to[isource,l=$c+a\hat e_i$] (-4.8,1.4) to[nos,l=$\phi_2$,color=idl] (-4.8,-1.3);
% ---- summing bus ----
\draw (-6.3,-1.3) -- (1.6,-1.3);
\fill (-3.4,-1.3) circle (1.6pt);
\node[above] at (-3.75,-1.3) {$A_i$};
% ---- latch (kill) switch to Vref ----
\draw[idl] (-3.4,-1.3) to[nos,l_=$S_L$,color=idl] (-3.4,-3.9);
% ---- M1 diode ----
\node[nmos] (M1) at (-1.6,-2.6) {};
\draw (M1.drain) -- (-1.6,-1.3);
\draw (M1.gate) -- ++(-0.45,0) |- (-1.6,-1.75);
\draw (M1.source) -- (-1.6,-3.9);
\node at ($(M1)+(0.75,0.1)$) {$M_1$};
% ---- comparator + latch ----
\node[blk, minimum width=2.1cm, minimum height=0.9cm] (cmp) at (-1.15,0.85) {comparator\\+ latch $a_i$};
\fill (-1.15,-1.3) circle (1.6pt);
\draw[idl,-{Latex[length=1.6mm]}] (-1.15,-1.3) -- node[right,font=\large]{$V_A$} (cmp.south);
\draw[idl, dashed] (cmp.west) -| (-3.1,-2.5);
\node[idl, font=\large] at (-2.85,1.25) {$\overline{a_i}$};
% ---- incoming weights ----
\draw[-{Latex[length=1.8mm]}, tblue] (0.8,0.1) -- (0.8,-1.2);
\node[above, tblue, font=\large, align=center] at (0.8,0.1) {weights\\in, (b)};
% ---- M2 with cascode ----
\node[nmos] (M2) at (2.6,1.5) {};
\draw (1.6,-1.3) |- (M2.gate);
\node[nmos] (MK) at (2.6,3.1) {};
\draw (M2.drain) -- (MK.source);
\draw (MK.drain) -- (2.6,4.1); \draw (2.3,4.1) -- (2.9,4.1);
\draw (MK.gate) -- ++(-0.3,0) node[left]{$V_{\rm cas}$};
\node at ($(M2)+(0.75,0.1)$) {$M_2$};
\node at ($(MK)+(0.75,0.1)$) {$M_K$};
% ---- state node ----
\draw (M2.source) -- (2.6,0) -- (8.6,0);
\fill (2.6,0) circle (1.6pt); \node[below left] at (2.6,0) {$C_i$};
\draw (3.7,0) to[C,l_=$C$] (3.7,-2.2) node[ground]{};
\draw (5.0,0) to[isource,l=$2I_0$] (5.0,-2.2) node[ground]{};
\draw[-{Latex[length=1.8mm]}, tblue] (6.7,0) -- (6.7,-1.25);
\node[below, tblue, font=\large, align=center] at (6.9,-1.3) {to $M_d$ of\\out-weights};
% ---- M3 ----
\node[nmos] (M3) at (8.6,1.5) {};
\draw (M3.source) -- (8.6,0);
\draw (M3.drain) -- (8.6,2.5);
\draw (8.6,4.1) to[isource,l_=$I_0$] (8.6,2.5);
\draw (8.3,4.1) -- (8.9,4.1);
\draw (M3.gate) -- ++(-0.35,0) |- (8.6,2.25);
\fill (8.6,2.25) circle (1.6pt);
\node at ($(M3)+(0.75,0.1)$) {$M_3$};
% ---- V_B bus ----
\draw (8.6,2.25) -- (14.0,2.25); \node[above] at (9.5,2.25) {$V_B$};
% ---- M4 and read switch ----
\node[nmos] (M4) at (11.2,-2.6) {};
\draw (M4.gate) -- ++(-0.35,0) coordinate (g4);
\draw (g4) -- (g4 |- 0,2.25);
\fill (g4 |- 0,2.25) circle (1.6pt);
\draw (M4.source) -- (11.2,-3.9);
\draw (M4.drain) to[nos,l_=$\phi_{\rm rd}$,color=idl] (11.2,0.6);
\draw[-{Latex[length=1.8mm]}] (11.2,0.6) -- (12.0,0.6) node[right]{$x_i$ out};
\node at ($(M4)+(0.75,0.1)$) {$M_4$};
% ---- sample and hold of V_B ----
\draw[idl] (14.0,2.25) to[nos,l=$\phi_h$,color=idl] (14.0,0.7);
\draw[idl] (14.0,0.7) to[C,l_=$C_h$,color=idl] (14.0,-1.0) node[ground]{};
\draw[idl,-{Latex[length=1.6mm]}] (14.0,0.7) -- (15.2,0.7);
\node[idl, font=\large, align=center] at (15.85,0.05) {held $x_i$,\\to update};
%=========================== (b) reciprocal weight =========================
\begin{scope}[shift={(18.4,0)}]
\node[anchor=west] at (-1.6,4.6) {\textbf{(b)} reciprocal weight $i\!\leftarrow\!j$};
\draw (-1.6,-3.9) -- (7.4,-3.9); \node[right] at (7.45,-3.9) {$V_{\rm ref}$};
% sender state node and Md
\fill (0,-1.4) circle (1.6pt); \node[above left] at (-0.45,-1.4) {$C_j$};
\draw[tblue] (-1.4,-1.4) -- (0,-1.4);
\draw (0,-1.4) to[isource,l_=$I_w$] (0,-3.5) node[ground]{};
\node[nmos] (Md) at (0,-0.3) {};
\draw (Md.source) -- (0,-1.4);
\draw (Md.drain) -- (0,1.0);
\draw (0,3.2) to[isource,l_=$I_w$] (0,1.0);
\draw (-0.3,3.2) -- (0.3,3.2);
\draw (Md.gate) -- ++(-0.35,0) |- (0,0.75);
\fill (0,0.75) circle (1.6pt); \node[above right] at (0,0.75) {$D$};
\node at ($(Md)+(0.7,0.1)$) {$M_d$};
% Mw
\node[nmos] (Mw) at (2.6,-2.6) {};
\draw (Mw.gate) -- ++(-0.35,0) coordinate (gw);
\draw (0,0.75) -| (gw);
\draw (Mw.source) -- (2.6,-3.9);
\draw (Mw.drain) -- (2.6,-0.6);
\fill (2.6,-0.6) circle (1.6pt);
\node at ($(Mw)+(0.75,0.1)$) {$M_w$};
% negative path
\draw[torange,-{Latex[length=1.8mm]}] (2.6,-0.6) -- (4.4,-0.6);
\node[right, torange, font=\large] at (4.4,-0.6) {$A_i$, $W_{ij}<0$};
% positive path: PMOS mirror
\draw[tblue] (2.6,-0.6) -- (2.6,1.0);
\node[pmos, tblue, xscale=-1] (Mpa) at (2.6,1.9) {};
\node[pmos, tblue] (Mpb) at (5.0,1.9) {};
\draw[tblue] (Mpa.drain) -- (2.6,1.0);
\draw[tblue] (Mpa.source) -- (2.6,3.2); \draw[tblue] (2.3,3.2) -- (2.9,3.2);
\draw[tblue] (Mpb.source) -- (5.0,3.2); \draw[tblue] (4.7,3.2) -- (5.3,3.2);
\draw[tblue] (Mpa.gate) -- (Mpb.gate);
\draw[tblue] (3.8,1.9) |- (2.6,1.15);
\fill[tblue] (2.6,1.15) circle (1.6pt);
\fill[tblue] (3.8,1.9) circle (1.6pt);
\draw[tblue,-{Latex[length=1.8mm]}] (Mpb.drain) -- (5.0,0.4);
\node[below, tblue, font=\large] at (5.0,0.4) {$A_i$, $W_{ij}>0$};
\node[tblue] at ($(Mpa)+(-0.8,0.1)$) {$M_{pa}$};
\node[tblue] at ($(Mpb)+(0.8,0.1)$) {$M_{pb}$};
\node[font=\large, align=left, anchor=west] at (3.8,-1.75) {$I_{\rm out}=\frac{I_w}{I_0}\,x_j$\\[1pt] sign bit selects path};
\end{scope}
\end{circuitikz}